\documentclass[twocolumn,showpacs,amsmath,amssymb,aps,prd,floats]{revtex4-1}
\usepackage{bm}% bold math
\usepackage{graphicx}
\usepackage{times}
\usepackage{dcolumn}% Align table columns on decimal point
\newcommand{\EQ}{\begin{equation}}
\newcommand{\EN}{\end{equation}}
\newcommand{\EQA}{\begin{eqnarray}}
\newcommand{\ENA}{\end{eqnarray}}

\newcommand{\Eq}[1]{Eq.~(\ref{#1})}

\newcommand{\xx}{\mbox{\boldmath $x$} {}}
\newcommand{\nn}{\mbox{\boldmath $n$} {}}
\newcommand{\kk}{\mbox{\boldmath $k$} {}}
\newcommand{\ssvec}{\mbox{\boldmath $s$} {}}

\newcommand{\kkk}{\hat{\bm{k}}}
\newcommand{\sss}{\hat{\bm{s}}}

\begin{document}

\title{ 
Primordial Magnetic Field Limits from Cosmic Microwave Background Bispectrum \\ of Magnetic Passive Scalar Modes }
\author{Pranjal Trivedi}
\affiliation{Sri Venkateswara College, University of Delhi, Delhi 110021, India \\ \& Department of Physics and Astrophysics, 
University of Delhi, Delhi 110007, India.}
\email{ptrivedi@physics.du.ac.in}
\author{Kandaswamy Subramanian}
\affiliation{IUCAA, Post Bag 4, Ganeshkhind, Pune 411 007, India.}
\email{kandu@iucaa.ernet.in}
\author{T. R. Seshadri}
\affiliation{Department of Physics and Astrophysics, University of Delhi,
Delhi 110007, India.}
\email{trs@physics.du.ac.in}

\date{\today}

\begin{abstract}
Primordial magnetic fields lead to non-Gaussian signals in the cosmic microwave
background (CMB) even at the lowest order, as magnetic stresses and the
temperature anisotropy they induce depend quadratically on the magnetic field. 
In contrast, CMB non-Gaussianity due to inflationary scalar perturbations arises
only as a higher order effect.  Apart from a compensated scalar mode, stochastic
primordial magnetic fields also produce scalar anisotropic stress that remains
uncompensated till neutrino decoupling. This gives rise to an adiabatic-like
scalar perturbation mode that evolves passively thereafter (called
the passive mode). We compute the CMB reduced bispectrum ($b_{l_{_1}l_{_2}l_{_3}}$) induced
by this passive mode, sourced via the Sachs-Wolfe effect, on large angular scales.
For any configuration of bispectrum, taking a partial sum over mode-coupling terms, 
we find a typical value of $l_1(l_1+1)l_3(l_3+1)
b_{l_{_1}l_{_2}l_{_3}} \sim 6-9 \times 10^{-16}$, for a magnetic field of $B_0
\sim 3$ nG, assuming a nearly scale-invariant magnetic spectrum . We also
evaluate, in full, the bispectrum for the squeezed collinear configuration over
all angular mode-coupling terms and find  $l_1(l_1+1)l_3(l_3+1)
b_{l_{_1}l_{_2}l_{_3}} \approx -1.4 \times 10^{-16}$. These values are more than $\sim 10^6$
times larger than the previously calculated magnetic compensated scalar
mode CMB bispectrum. Observational limits on the bispectrum from WMAP7 data
allow us to set upper limits of $B_0 \sim 2$ nG on the present value of the
cosmic magnetic field of primordial origin. This is over 10 times more
stringent than earlier limits on $B_0$ based on the compensated mode bispectrum.

\end{abstract}


\maketitle

%*=*=*=*=*=*=*=*=*=*=*
\section{Introduction}
The origin and evolution of large-scale magnetic fields in the universe
is not yet clearly understood. There is observational evidence for micro-Gauss
strength magnetic fields, ordered on kiloparsec to 10 kiloparsec scales, in galaxies and clusters of galaxies  
\cite{beck09,vogt_ensslin03,BS05,kulsrud_zweibel08,beck_ARAA}. There is also evidence for galactic scale fields 
at higher redshift \cite{bernet08,kronberg08}.
There are only tentative indications, to date, of fields not associated with individual 
galaxies or clusters \cite{kim89}. 
Further, recent $\gamma$-ray observations using {\it Fermi}/LAT data provide hints of a detection \cite{ando10} 
or a lower bound \cite{neronov10} of $B \sim 10^{-16}$ G on intergalactic scales. Constraints on purely cosmological magnetic 
fields have also been derived from the CMB \cite{kandu_AN06,durrer_NAR07,barrow97,SB98PRL,durrer00,trsks01,mack02,lewis04,
SS05,gopal_sethi05,kahn05,giovannini06,giovan_kunze08,yam08,finelli08,pao09,giovan09,yam10,pao10,SL10}, big bang nucleosynthesis 
and polarized radiation from extragalactic radio sources (for a review see e.g. \cite{widrow02}).

It is generally accepted that the observed fields require a seed magnetic field,
possibly of primordial origin, which are then 
amplified by astrophysical processes \cite{BS05,kulsrud_zweibel08,widrow02,moffatt,krause,ruzmaikin}. 
A strong enough seed may mainly require amplification due to flux-freezing which arises during the collapse to
form structures \cite{kulsrud_zweibel08}. 
On the other hand a weak seed would require considerable dynamo action as well 
\cite{BS05,kulsrud_zweibel08,moffatt,krause,ruzmaikin}. 
As yet no compelling theory exists for the origin of strong  
primordial fields. Equally, dynamo theories are not without 
their own difficulties, in particular, how large-scale dynamos 
lead to coherent enough fields on saturation \cite{BS05}.
Thus it is important to keep an open mind on the issue
of the origin of magnetic fields.

Large-scale primordial fields could 
have resulted from phase transitions in the early universe, for example during
the inflationary era \cite{TW88,ratra92,giovannini_string07,martin_yoko08,kandu_AN10}. An important way to constrain or detect 
such fields is via their imprints on the Cosmic Microwave Background (CMB) anisotropies.
Considerable work has already been done to calculate the magnetic field
signals in the power spectrum (both temperature and polarization) 
of the CMB anisotropies \cite{barrow97,SB98PRL,durrer00,trsks01,mack02,lewis04,SS05,gopal_sethi05,kahn05,giovannini06,giovan_kunze08,
yam08,finelli08,pao09,giovan09,yam10,pao10,SL10,SL09,brown10,chen04,naselksy04,bernui08,kahn08}. The possibility of non-Gaussian 
signals in the CMB temperature anisotropies 
has drawn increased attention of late. This possibility is particularly relevant when one considers magnetically induced anisotropies 
in the CMB, for the following reason: In standard inflationary models, the Gaussian statistics of the quantum fluctuations in the 
inflaton field leads to Gaussian statistics for the CMB temperature field. In such models, any non-Gaussian signal in the temperature 
field arises generically due to higher order effects \cite{wang_kamion00,komatsu_spergel01,bartolo04,FS07,riotto08,
yadav_wandelt10}. 

The situation is different in the case of magnetically induced signals. 
Magnetic stresses depend quadratically on the field. Hence, even if the fields are assumed to be Gaussian, the corresponding 
stresses are not. This implies that the CMB temperature anisotropy, which these stresses induce, do not have Gaussian statistics. 
Unlike the case of CMB non-Gaussianity from inflationary perturbations, in the case of magnetic fields, the CMB non-Gaussianity is 
a leading order effect. Thus the study of CMB non-Gaussianity has a special significance in the context of probing and detecting 
primordial cosmological magnetic fields.

Preliminary studies of such non-Gaussian signals in the CMB induced by primordial magnetic fields have begun 
\cite{SS09,caprini09,cai10}, 
based on earlier calculations of non-Gaussianity in the magnetic stress energy \cite{BC05}. 
These have been limited to the scalar mode signals on large scales and further 
restricted to a component referred to as the compensated scalar mode by \cite{SL09}.  However, the work of 
Shaw and Lewis \cite{SL09} has revealed that much larger CMB anisotropies can result from scalar perturbations sourced
by the magnetic anisotropic stresses prior to neutrino decoupling (also see \cite{kojima10,bonvin10,kunze10}). 
This mode is referred to as the passive scalar mode \cite{SL09}. In this paper we calculate the contribution 
to the CMB bispectrum from the passive scalar mode. As we show here, this contribution greatly dominates 
over the contribution calculated earlier by two of us \cite{SS09} (hereafter referred to as SS09), 
and also calculated by \cite{caprini09}. The current work allows tighter limits to be placed on 
primordial magnetic fields. 

In the next section we describe the properties of the primordial stochastic
magnetic field and in section 3 we describe the perturbation induced due to the
passive scalar mode. The magnetic CMB temperature anisotropy is described in Section 4 and
in its subsections we calculate the three-point correlation of the scalar
anisotropic stress in two different configurations. In section 5 we calculate
the passive mode reduced bispectrum for the different configurations and in section 6 we use
the reduced magnetic bispectra to put constraints on the primordial magnetic
field. Our conclusions are summarized in section 7.

%*=*=*=*=*=*=*=*=*=*=*
\section{Primordial Stochastic Magnetic Field}

We consider a stochastic magnetic field $B$ which is a Gaussian random field characterized and completely specified by its 
energy power spectrum $M(k)$. In addition we assume the magnetic field is non-helical.
 
On galactic and larger scales, any velocity induced by Lorentz forces is generally so small that it does not lead to 
appreciable distortion of the initial field \cite{jedam98,SB98}.  Hence, the magnetic field simply redshifts away 
as ${\bf B}({\bf x},t)={\bf b}_{0}({\bf x})/a^{2}$, where, ${\bf b}_{0}$ is the magnetic field at the present epoch 
(i.e. at $z=0$ or $a=1$).

We define ${\bf b} ({\bf k})$ as the Fourier transform of the magnetic field ${\bf b}_0 ({\bf x})$. 
The energy power spectrum is defined by the relation 
$\langle b_{i}({\bf k})b^*_{j}({\bf q})\rangle=(2\pi)^3 \delta({{\bf k}-{\bf q}})P_{ij}({\bf k})M(k)$, 
where $P_{ij}({\bf k}) = (\delta_{ij} - k_ik_j/k^2)$ is the projection operator ensuring ${\bf \nabla}\cdot{\bf b_0} =0$. 
This gives $\langle {\bf b}_{0}^{2} \rangle=2\int (dk/k)\Delta _{b}^{2}(k)$, where $\Delta _{b}^{2}(k)=k^{3}M(k)/(2\pi ^{2})$ 
is the power per logarithmic interval in $k$-space residing in the stochastic magnetic field.

As in \cite{SSB03}, we assume a power-law magnetic energy spectra,
$M(k)=Ak^{n}$ that has a cutoff at $k=k_{c}$,
where $k_{c}$ is the Alfv\'{e}n-wave damping length-scale
\cite{jedam98,SB98}. We fix $A$ by demanding
that the variance of the magnetic field smoothed over a
`galactic' scale, $k_{G}=1h {\rm Mpc}^{-1}$, (using a sharp $k$-space
filter) is $B_{0}$. This gives, (for $n>-3$ and for $k<k_c$)
\EQ
\Delta _{b}^{2}(k)= \frac{k^3M(k)}{2\pi^2}
=\frac{B_0^2}{2}(n+3)\left(\frac{k}{k_{G}}\right)^{3+n}.
\EN

The magnetic spectral index is restricted to $-3 \lesssim n < -3/2$. Blue
spectral indices are strongly disfavoured by many observations like the CMB
power spectra \cite{yam10}, \cite{pao10} and especially by the gravitational
wave limits of \cite{cap02}.

%*=*=*=*=*=*=*=*=*=*=*
\section{Passive Scalar Mode}
The stress tensor (space-space part of the 
energy-momentum tensor) for magnetic fields in terms of the present day magnetic field value $ {\bf b_0}$ is
\EQ
T^i_j({\bf x}) = \frac{1}{4 \pi a^4} \left( \frac{1}{2} b_0^2({\bf x}) \delta^i_j - b_0^i({\bf x}) {b_0}_j({\bf x}) \right)
\label{Tij_b_nought}
\EN
In Fourier space, the product of magnetic fields becomes a convolution
\EQ
S^i_j({\bf k}) = \int {b}^i({\bf q}) {b}_j({\bf k-q}) d^3 {\bf q}
\label{Tij_convolution}
\EN
\EQ
T^i_j({\bf k}) = \frac{1}{4 \pi a^4} \left( \frac{1}{2} S^{\alpha}_{\alpha}({\bf k}) \delta^i_j - S^i_j({\bf k}) \right).
\label{Tij_b_nought_fourier}
\EN
This can be expressed in terms of the magnetic perturbations to the energy-momentum tensor as
\EQ
T^i_j({\bf k}) = p_{\gamma} \left( \Delta_B \delta^i_j + {\Pi_B}^i_j \right)
\label{Tij_perturb}
\EN
where $\Delta_B$ and ${\Pi_B}^i_j$ are the perturbations in the energy density and anisotropic stress, respectively, as 
defined in \cite{SL09} and $p_{\gamma}$ is the radiation pressure.

The magnetic stresses are non-linear in the field but we assume that they are
always small compared to the total energy density and pressure of the photons,
baryons etc. Thus allowing a purely linear treatment of the perturbations. Hence
the scalar, vector and tensor perturbations decouple and evolve independently
and here we focus on the anisotropic scalar perturbations $\Pi_B(\kk)$ which are
given by applying the relevant projection operator to $T^i_j({\bf k})$
\cite{BC05}.
\EQ
\Pi_B(\kk) = -\frac{3}{2} \left( \kkk _i \kkk_j - \frac{1}{3} \delta_{ij} \right) \Pi_B^{ij}
\EN
Note that $\Pi_B(\kk)$ of \cite{SL09} is equal to $ -\tau^S(\kk)$ of \cite{BC05}.

Such an anisotropic stress term also arises due to neutrinos after they decouple. The net anisotropic stress tensor 
acts as the source term in the scalar Einstein's equations that lead to the Bardeen equation for the potential. Prior to 
neutrino decoupling, the only source of anisotropic stress is the magnetic 
field. Once the neutrinos decouple, the anisotropic stress due to neutrinos also contributes but 
with an opposite sign to that of the magnetic field, thus compensating the contribution from the magnetic field \cite{lewis04}. 
After compensation there are two modes of perturbation \cite{SL09}. The first one, called the passive mode is an adiabatic-like 
mode but has non-Gaussian statistics. It grows logarithmically in amplitude between the epochs of magnetic field generation and 
neutrino decoupling but then evolves passively after neutrino decoupling. This behaviour has also been confirmed in \cite{bonvin10} 
in the context of deriving the magnetic Sachs-Wolfe effect for a causally generated primordial magnetic field. The second, more 
well-studied perturbation \cite{giovan_kunze08,yam08,finelli08,pao09}, is called the compensated mode. 

The final curvature perturbation due to the passive mode is given by equation (86) of Shaw \& Lewis \cite{SL09},
\EQ
\zeta=\zeta\left( \tau_B\right)  - \frac{1}{3} R_{\gamma} \Pi_B \left[ \ln \left( \frac{\tau_{\nu}}{\tau_B}\right) + 
\left( \frac{5}{8R_{\nu}} - 1\right) \right].
\EN
in the conformal Newtonian gauge. The evolution of the curvature perturbation has also been discussed (in synchronous gauge) 
in \cite{kojima10} for the case of an extra source of anisotropic stress cancelling the neutrino anisotropic stress. The role of 
anisotropic stresses on CMB has also been discussed by \cite{giovannini10_spectator}. Here $R_\nu \sim 0.4$ is the fractional 
contribution of neutrino energy density towards the total energy density of the relativistic component. The fractional 
contribution of radiation energy density is $R_{\gamma} = 1 - R_\nu \sim 0.6$. The $\zeta\left( \tau_B\right)$ 
term represents all primordial contributions to the curvature perturbation. The log term shows the growth in the curvature 
between the epochs of magnetic field generation $\tau_B$ and neutrino decoupling $ \tau_{\nu}$. This log term is always bigger 
than 10 for different choices of $\tau_B$ so we can neglect the $((5/8R_{\nu}) - 1)$ term with less than $5 \%$ error. After 
neutrino decoupling, the anisotropic stress $\Pi_B$ is compensated and there is a remnant adiabatic mode with amplitude given 
by \cite{SL09}.
\EQ
\zeta \sim  - \frac{1}{3} R_{\gamma} \Pi_B \ln \left( \frac{\tau_{\nu}}{\tau_B}\right).
\label{zeta}
\EN
This so-called passive mode evolves passively like an adiabatic perturbation but
the statistics of $\Pi_B$ are non-Gaussian unlike the standard inflationary
adiabatic mode. Also, unlike the compensated mode, the passive mode amplitude
depends logarithmically on the epoch of magnetic field generation. Therefore,
limits on the primordial magnetic field strength arising from this passive mode
will be somewhat sensitive to the assumed model and epoch of magnetic field
generation.

%*=*=*=*=*=*=*=*=*=*=*
\section{Magnetic CMB Anisotropy and Three-point Correlation}

We consider the CMB temperature anisotropies that are sourced by the scalar
passive mode. On large angular scales the important contribution is via the
magnetically induced Sachs-Wolfe effect
\EQ
\frac{\Delta T}{T}(\nn) = \frac{1}{3} \, \Phi(\xx_0 -\nn D^*) = \frac{1}{5} \, \zeta(\xx_0 -\nn D^*)
\EN
where the second equality is from, for example, Eq. (10.42) of \cite{lythliddle}.
The unit vector ${\bf n}$ gives the direction of observation and $D^*$ is the
(angular diameter) distance to the surface of last scatter. Then employing Eq.
(\ref{zeta}) we get
\EQA
\frac{\Delta T}{T}(\nn) &=&  -\frac{1}{15} R_{\gamma} \ln \left( \frac{\tau_{\nu}}{\tau_B}\right) 
\Pi_B(\xx_0 -\nn D^*) \nonumber \\ &=& 
{\cal R}_{p} \Pi_B(\xx_0 -\nn D^*).
\label{delta_t}
\ENA 
where we define
\EQ
{\cal R}_{p} = -\frac{1}{15} R_{\gamma} \ln \left( \frac{\tau_{\nu}}{\tau_B}\right).
\EN

In SS09 we had calculated the bispectrum due to the compensated scalar mode 
for which 
$\Delta T/T = (R_{\gamma}/20) \Delta_B$ 
\cite{giovannini07_PMC}. Equation (\ref{delta_t}) for the 
passive scalar mode leads to a 
$\Delta T/T$ 
which is larger by a factor 
${\cal A} = (4/3) \ln \left( \tau_{\nu}/\tau_B \right) (\Pi_B / \Delta_B)$. 
Assuming that 
the fractional perturbations $\Delta_B$ and $\Pi_B$ are of the same order, we get ${\cal A}$ of order 50 for $\tau_B \approx 
10^{14}$ GeV. 
As we will see below for the magnetic spectra we consider, this alone leads to an enhanced passive scalar mode contribution to the 
bispectrum of order ${\cal A}^3 \approx 10^5$ and thus stronger limits on the primordial magnetic field by a factor ${\cal A}^{0.5}$.
We assume instantaneous recombination which is a good approximation for large angular scales. There could also be additional 
integrated Sachs-Wolfe (ISW) contributions to 
$\Delta T/T(\nn)$.

The CMB temperature fluctuation in a direction $(\nn)$ at the observer can be expanded in terms of the spherical harmonics to give 
\EQ
\frac{\Delta T(\nn)}{T} = \sum_{l m} a_{lm} Y_{lm}(\nn).
\EN
where, for the passive scalar mode,
\EQ
a_{lm}= 4\pi (-i)^l \int \frac{d^3 k}{(2\pi)^3} ~
{\cal R}_p ~ \Pi_B(\kk) ~ j_l(kD^*) Y^*_{lm}(\hat \kk). 
\label{alm}
\EN
Here $\Pi_B(\kk)$ is the Fourier transform of $\Pi_B({\bf x})$.

The non-Gaussianity in the CMB temperature anisotropy can be evaluated by
calculating its 3-point correlation function (in harmonic space), called the
bispectrum, $B^{m_{_1}m_{_2}m_{_3}}_{l_{_1}l_{_2}l_{_3}}$. We assume that the
magnetic perturbations are uncorrelated with the primary (inflationary)
perturbations in the CMB.

In terms of the ${a_{lm}}$'s the CMB bispectrum is given by,
\EQ
B^{m_{_1}m_{_2}m_{_3}}_{l_{_1}l_{_2}l_{_3}}=\langle a_{{l_1}{m_1}}a_{{l_2}{m_2}}a_{{l_3}{m_3}} \rangle.
\EN
From Eq.~(\ref{alm}) we can express $B^{m_{_1}m_{_2}m_{_3}}_{l_{_1}l_{_2}l_{_3}}$ as
\EQ
B^{m_{_1}m_{_2}m_{_3}}_{l_{_1}l_{_2}l_{_3}}
= {\cal R}_p^3\int
\left[\prod_{i=1}^3 (-i)^{l_i}
\frac{d^3k_i}{2\pi^2} j_{_{l_i}}(k_{_i}D^*)Y^*_{l_im_i}(\hat{\kk}_{_i})\right]
\zeta_{_{123}}
\label{bispec}
\EN
with $\zeta_{_{123}}$ defined as,
\EQ
\zeta_{_{123}}=\langle \hat\Pi_B(\kk_{_1})\hat\Pi_B(\kk_{_2})
\hat\Pi_B(\kk_{_3}) \rangle.
\EN
The function $\zeta_{_{123}}$ is the three-point correlation of $\hat\Pi_B(\kk)$
and involves a $6$-point correlation function of the magnetic fields. Using
Wick's theorem it can be written as a function of magnetic spectra in an
analogous manner to \cite{BC05} and Eq. (7) of SS09, $\zeta_{_{123}} =
\delta(\kk_1 + \kk_2 + \kk_3) ~ \psi_{123}$, where
\EQA
\psi_{123} = \frac{1}{(4\pi p_\gamma)^3}
\int &d^3s& M(\vert \kk_1 + \ssvec   \vert) M(s) M(\vert \ssvec   -\kk_3 \vert) \nonumber \\
&\times& (\mathcal F_{\Pi_B\Pi_B\Pi_B}).
\label{psi}
\ENA
where $\mathcal F_{\Pi_B\Pi_B\Pi_B}$ is the angular component of the three-point
correlation $\zeta_{_{123}}$ of the scalar anisotropic stress $\Pi_B(\kk)$ in
mode-coupling integral $\psi_{123}$.

%*=*=*=*=*=*=*=*=*=*=*
\subsection{Three-Point Correlation of Scalar Anisotropic Stress}
This angular component $\mathcal F_{\Pi_B\Pi_B\Pi_B}$ is given by a 58-term
expression in \cite{BC05} derived by applying relevant projection operators to
extract the scalar part of the full bispectrum i.e the six-point correlation of
the magnetic fields. In this particular case, the operator is given by $\mathcal
A_{ijklmn} = (-1)^3 \mathcal Q_{ij}(\kk_3) \mathcal Q_{kl}(\kk_1) \mathcal
Q_{mn}(\kk_2)$ where $\mathcal Q_{ab}(\kk) \equiv \delta_{ab} -(3/2)P_{ab}(\kk)$
and $P_{ab}(\kk) = \delta_{ab}-\kkk_a\kkk_b$ is the projection operator. We
present the full angular component (in our notation we have absorbed a factor of
8 multiplying $\mathcal F$ in \cite{BC05} into our definition of $\mathcal
F_{\Pi_B\Pi_B\Pi_B}$):
\EQ
\mathcal{F}_{\Pi_B \Pi_B \Pi_B }=\sum_{n=0}^6\mathcal{F}_{\Pi_B \Pi_B \Pi_B }^{n}
\label{F_gen}
\EN
with \begin{widetext}
\EQA
 \mathcal{F}_{\Pi_B \Pi_B \Pi_B }^0&=&-9 \nonumber \\
 \mathcal{F}_{\Pi_B \Pi_B \Pi_B }^1&=&0 \nonumber \\
 \mathcal{F}_{\Pi_B \Pi_B \Pi_B }^2&=&\Big(
  \bar{\beta}^2+\bar{\gamma}^2+\bar{\mu}^2
  +9(\theta_{13}^2+\theta_{23}^2+\theta_{12}^2)
%   \nonumber \\ && \quad
  +3(\alpha_3 ^2+\alpha_1 ^2+\alpha_2 ^2+\beta_3 ^2+\beta_1 ^2+\beta_2 ^2+\gamma_3 ^2+\gamma_1 ^2+\gamma_2 ^2)
  \Big) \nonumber \\
 \mathcal{F}_{\Pi_B \Pi_B \Pi_B }^3&=&-3\bigg(
  \bar{\mu}(\beta_3 \gamma_3 +\beta_1 \gamma_1 +\beta_2 \gamma_2 +\frac{1}{3}\bar{\beta}\bar{\gamma})
  +\bar{\gamma}(\alpha_3 \gamma_3 +\alpha_1 \gamma_1 +\alpha_2 \gamma_2 )
%   \nonumber \\ && \quad
  +\bar{\beta}(\alpha_3 \beta_3 +\alpha_1 \beta_1 +\alpha_2 \beta_2 )
   \nonumber \\ && \quad
  +3\theta_{13}(\alpha_3 \alpha_1 +\beta_3 \beta_1 +\gamma_3 \gamma_1 )
%   \nonumber \\ && \quad
  +3\theta_{23}(\alpha_3 \alpha_2 +\beta_3 \beta_2 +\gamma_3 \gamma_2 )
  +3\theta_{12}(\alpha_1 \alpha_2 +\beta_1 \beta_2 +\gamma_1 \gamma_2 )
   \nonumber \\ && \quad
  +9\theta_{13}\theta_{23}\theta_{12}
  \bigg) \nonumber \\
 \mathcal{F}_{\Pi_B \Pi_B \Pi_B }^4&=&3\bigg(
  \bar{\gamma}\bar{\mu}\alpha_3 \beta_3 +\bar{\beta}\bar{\mu}\alpha_1 \gamma_1 +\bar{\beta}\bar{\gamma}\beta_2 \gamma_2 
%    \nonumber \\ && \quad
  +3\big(\bar{\mu}\theta_{13}\beta_3 \gamma_1 +\bar{\gamma}\theta_{23}\alpha_3 \gamma_2 +\bar{\beta}\theta_{12}\alpha_1 \beta_2 \big)
    \nonumber \\ && \quad
  +3\big(\alpha_3 \beta_3 (\alpha_1 \beta_1 +\alpha_2 \beta_2 )
  +\alpha_1 \gamma_1 (\alpha_3 \gamma_3 +\alpha_2 \gamma_2 )
  +\beta_2 \gamma_2 (\beta_3 \gamma_3 +\beta_1 \gamma_1 )\big)
    \nonumber \\ && \quad
  +9(\theta_{13}\theta_{23}\gamma_1 \gamma_2 +\theta_{13}\theta_{12}\beta_3 \beta_2 +\theta_{23}\theta_{12}\alpha_3 \alpha_1 )
  \bigg) \nonumber \\
% \nonumber \\
%\ENA
%\EQA
 \mathcal{F}_{\Pi_B \Pi_B \Pi_B }^5&=&-9\bigg(
  \bar{\mu}\alpha_3 \beta_3 \alpha_1 \gamma_1 
  +\bar{\gamma}\alpha_3 \beta_3 \beta_2 \gamma_2 +\bar{\beta}\alpha_1 \gamma_1 \beta_2 \gamma_2 
%   \nonumber \\ && \quad
  +3(\theta_{13}\beta_3 \gamma_1 \beta_2 \gamma_2 +\theta_{23}\alpha_3 \alpha_1 \gamma_1 \gamma_2 +\theta_{12}\alpha_3 \beta_3 \alpha_1 \beta_2 )
  \bigg) \nonumber \\
 \mathcal{F}_{\Pi_B \Pi_B \Pi_B }^6&=&27\alpha_3 \beta_3 \alpha_1 \gamma_1 \beta_2 \gamma_2.
\label{58terms}
\ENA
\end{widetext}
where
\EQ
\bar{\beta} = (\widehat{\ssvec  }\cdot\widehat{\kk_3 - \ssvec  }), \, 
\bar{\gamma} = (\widehat{\ssvec  }\cdot\widehat{\kk_1 + \ssvec  }), \, 
\bar{\mu} = (\widehat{\kk_3 - \ssvec  }\cdot\widehat{\kk_1 + \ssvec  }).
\EN
where the hat on a vector denotes its unit vector. \\
Also,
\EQ
\alpha_a = \kkk_a \cdot \ssvec  , \quad
\beta_a = \kkk_a \cdot \widehat{\kk_3 - \ssvec  }, \quad
\gamma_a = \kkk_a \cdot \widehat{\kk_1 + \ssvec  }
\EN
and 
\EQ
\theta_{ab} = \kkk_a \cdot \kkk_b.
\EN
where our angle definitions are consistent with \cite{BC05} and slightly different to SS09.

For simplicity, we evaluate this $\mathcal F_{\Pi_B \Pi_B \Pi_B }$ expression to find the bispectrum in the following two cases below. 
We expect that the magnitude of the bispectrum will be of similar order for a general case.

\underline{Case I} - Consider any bispectrum configuration but include only the $s$-independent terms (constant for any particular 
configuration) in the $\mathcal F_{\Pi_B \Pi_B \Pi_B }$ expression.  Then the $s$-integral in Eq. (\ref{psi}) is 
performed without reference to any particular bispectrum configuration.

\underline{Case II} - The squeezed collinear configuration where we calculate fully the mode-coupling  integral over all 
angular terms in the $\mathcal F_{\Pi_B \Pi_B \Pi_B }$ expression.

\subsection{Case I: Evaluation with $s$-independent Terms in $\mathcal F_{\Pi_B \Pi_B \Pi_B }$}

The $s$-independent terms include  $\theta_{12},\theta_{23},\theta_{13}$ that are constant for a given $(\kk_1,\kk_2,\kk_3)$ 
configuration. For each configuration considered, we calculate the sum of the five $s$-independent terms in Eq. (\ref{58terms}) for 
$\mathcal F_{\Pi_B \Pi_B \Pi_B }$ 
\EQ
\mathcal F_{\Pi_B \Pi_B \Pi_B }^{\text{I}} = -9 + 9 \left( \theta_{12}^2 + \theta_{23}^2 + \theta_{13}^2 \right) -27 \left( \theta_{13}\theta_{12}\theta_{23}\right).
\label{F_const}
\EN
We give in Table I the values of $m = \mathcal F_{\Pi_B \Pi_B \Pi_B }^{\text{I}} (\kk_1,\kk_2,\kk_3) $ for specific 
configurations $(\kk_1,\kk_2,\kk_3)$. Note that $m$ happens to vanish exactly for the local isosceles configuration.
%=========================TABLE===========================
\begin{table}
\caption{The sum of $s$-independent terms $m = \mathcal F_{\Pi_B \Pi_B
\Pi_B}^{\text{I}}$ in four different configurations $(\kk_1,\kk_2,\kk_3)$ for
evaluating the Case I bispectrum.}
\begin{tabular}{cccc} 
\hline \hline \\
\multicolumn{1}{c}{Configuration} &
\multicolumn{1}{c}{$(\kk_1,\kk_2,\kk_3) $} &
\multicolumn{1}{c}{$(\theta_{12},\theta_{23},\theta_{13})$} &
\multicolumn{1}{c}{$m$}
 \\ [3 pt]
\hline \hline
%_/_/_/_/_/_/_/_/_/_/_/_/_/_/_/_/_/_/_/_/_/_/_/_/_/_/_/_/_/_/_/_/
\\
local &
$k_1 \sim k_3$  &
$(0,0,-1)$ &
$0$ \\
%_/_/_/_/_/_/_/_/_/_/_/_/_/_/_/_/_/_/_/_/_/_/_/_/_/_/_/_/_/_/_/_/
isosceles &
$k_2 \ll k_1, k_3 $ &
&
\\ [3 pt]
\hline
%_/_/_/_/_/_/_/_/_/_/_/_/_/_/_/_/_/_/_/_/_/_/_/_/_/_/_/_/_/_/_/_/
\\
equilateral &
$k_1 \sim k_2 \sim k_3$ &
$(\frac{1}{2},\frac{1}{2},\frac{1}{2})$ &
$-5.625$ \\ [3 pt]
\hline
%_/_/_/_/_/_/_/_/_/_/_/_/_/_/_/_/_/_/_/_/_/_/_/_/_/_/_/_/_/_/_/_/
\\
squeezed &
$k_1 \sim k_3$  &
$(1,-1,-1)$ &
$-8$ \footnotemark \\
%_/_/_/_/_/_/_/_/_/_/_/_/_/_/_/_/_/_/_/_/_/_/_/_/_/_/_/_/_/_/_/_/
collinear &
$k_2 \ll k_1, k_3 $  &
&
\\
%_/_/_/_/_/_/_/_/_/_/_/_/_/_/_/_/_/_/_/_/_/_/_/_/_/_/_/_/_/_/_/_/
&
$\kkk_1 = \kkk_2 = -\kkk_3$  &
&
\\ [3 pt]
\hline
%_/_/_/_/_/_/_/_/_/_/_/_/_/_/_/_/_/_/_/_/_/_/_/_/_/_/_/_/_/_/_/_/
\\
midpoint &
$k_1 \sim k_2 \sim \frac{k_3}{2}$ &
$(1,-1,-1)$ &
$-9$ \\
%_/_/_/_/_/_/_/_/_/_/_/_/_/_/_/_/_/_/_/_/_/_/_/_/_/_/_/_/_/_/_/_/
collinear&
$ \kkk_1 = \kkk_2 = -\kkk_3$ &
&
\\ [3 pt]
\hline \hline
\end{tabular}
\footnotetext[1]{For the squeezed collinear configuration case, $\mathcal
F_{\Pi_B \Pi_B \Pi_B }^{\text{I}}$ picks up another term $\bar{\mu}^2 \sim 1$}.
\end{table}
%==============================================================
Hence the Case I mode-coupling  integral reduces to 
\EQ
\psi_{123} = \frac{m}{(4\pi p_\gamma)^3} ~ \mathcal I = \frac{3^3~m}{(4\pi \rho_0)^3} ~ \mathcal I
\EN
where $\rho_0$ is the present day energy density of radiation and
\EQA
\mathcal I &=& \int d^3s M(\vert \kk_1 + \ssvec   \vert) M(s) M(\vert \ssvec   -\kk_3 \vert) \nonumber \\
&=& 2 \pi A^3 \int_{-1}^{1} d\mu \int_0^{\infty} ds s^{n+2}\left( s^2 + k_1^2 + 2sk_1 \nu \right)^\frac{n}{2} \nonumber \\
&\quad& \times \left( s^2 + k_3^2 - 2sk_3\mu \right)^\frac{n}{2} 
%\label{mode_I}
\ENA
where $\quad \nu = \kkk_1 \cdot \sss\quad $ and $\quad \mu = \kkk_3 \cdot \sss$.

We perform the mode-coupling integral using the technique discussed in \cite{trsks01,SSB03,mack02}. As $m$ vanishes for the
local isosceles configuration, $\psi_{123}$ is zero for this configuration. 
For the equilateral and squeezed collinear configurations we split the $s$ integral into %three
two sub-ranges $0<s<k_1 \sim k_3$ %$k_1<s<k_3$ 
and $s>k_1 \sim k_3$
We then approximate the integrands 
by assuming $s\ll k_1 \sim k_3$ 
and $s\gg k_1 \sim k_3$ for the respective sub-ranges.
For the midpoint collinear configuration we split the $s$ integral into two sub-ranges $0<s<k_1$ and $s> 2~k_1 \sim k_3$ 
while neglecting the very small contribution from the middle sub-range $k_1<s<2~k_1 \sim k_3 $. Again, we approximate the 
integrands by assuming $s\ll k_1$ and $s\gg 2~k_1 \sim k_3$ for the respective sub-ranges.
To derive numerical estimates of the bispectrum and magnetic field strengths we will focus on nearly scale-invariant spectra 
(which can be produced by an acausal mechanism like inflation), i.e $n \to -3$, which yield
\EQA
\mathcal I &\simeq& 4 \pi A^3\left[ \frac{k_1^{2n+3}k_3^n}{n+3} 
%+ \frac{k_3^{3n+3} -k_1^{2n+3}k_3^n}{2n+3}   NEGLECTING middle sub-range
- \frac{k_3^{3n+3}}{3n+3} \right] \nonumber \\
&\simeq& 4 \pi A^3  \frac{k_1^{2n+3}k_3^n}{n+3}
\ENA
The latter equation is obtained in the limit $n \to -3$ where we can neglect the
terms with 
%$(2n+3)^{-1}$ and 
$(3n+3)^{-1}$ compared to $(n+3)^{-1}$. Hence the
mode-coupling integral for Case I - taking only $s$-independent angular part - 
is
\EQ
\psi_{123} = \left( 4 \right)^4 m ~\frac{\pi^7}{k_G^6}
(n+3)^2 \left(\frac{k_1}{k_G}\right)^{2n+3} \left(\frac{k_3}{k_G}\right)^n V_A^6.
\label{psi_I}
\EN
Here we have defined $V_A$, the Alfv\'en velocity in the radiation dominated era as \cite{SB98}
\EQ
V_{A}={\frac{B_{0}}{(16\pi \rho _{0}/3)^{1/2}}}\approx 3.8\times
10^{-4}B_{-9},  
\label{alfvel}
\EN
with $B_{-9}\ \equiv (B_{0}/10^{-9}{\rm Gauss})$.\\

\subsection{Case II: Squeezed Collinear Configuration - All Angular Terms}

%------------CASE II-------------------
For Case II - We take the squeezed collinear bispectrum configuration as $k_1 \sim k_3$ and $k_2 \ll k_1, k_3$ and $\kkk_1 = \kkk_2 = -\kkk_3$. One wavevector ($\kk_2$) is negligibly small compared to the other two which are also in exactly opposite directions and of a similar magnitude. This affords considerable simplification and reduction of the angular terms given by Eq. (\ref{58terms}) in doing the mode-coupling integral. Using $\kk_3 \approx -\kk_1$ we see that
\EQ
\bar{\beta} = (\widehat{\ssvec  }\cdot\widehat{\kk_3 - \ssvec  }) \approx (\widehat{\ssvec  }\cdot\widehat{-\kk_1 - \ssvec  }) \approx -(\widehat{\ssvec  }\cdot\widehat{\kk_1 + \ssvec  }) \approx -\bar{\gamma}.
\EN
\EQA
\bar{\mu} &=& (\widehat{\kk_3 - \ssvec  }\cdot\widehat{\kk_1 + \ssvec  }) \approx (\widehat{ -\kk_1 - \ssvec   }\cdot\widehat{\kk_1 + \ssvec  }) 
\nonumber \\
&\approx& -(\widehat{\kk_1 + \ssvec  }\cdot\widehat{\kk_1 + \ssvec  }) \approx -1.
\ENA
\EQ
\alpha_1 = \alpha_2 = -\alpha_3 = \alpha = \nu = -\mu.
\EN
\EQ
\beta_1 = \beta_2 = -\beta_3 = \beta.
\EN
\EQ
\gamma_1 = \gamma_2 = -\gamma_3 = \gamma.
\EN
\EQ
\theta_{12} = 1,\, \theta_{23} = \theta_{13} = -1.
\EN
These relations substituted into Eq. (\ref{58terms}) reduce the 58 terms to a 10-term angular expression 
for the squeezed collinear configuration
\EQA
\mathcal{F}_{\Pi_B \Pi_B \Pi_B } &=& -8 + \bar{\beta}^2 + 9 \left( \mu^2 +2 \gamma^2  \right) + 6 \mu \bar{\beta} \gamma + 3 \bar{\beta}^2 \gamma^2 \nonumber \\ 
&\quad&- 9 \gamma^2 \left( 3\mu^2 + \gamma^2 \right) - 18 \mu \bar{\beta} \gamma^3 + 27 \mu^2 \gamma^4.
\label{F_sq_coll}
\ENA
We again perform the mode-coupling as discussed above and split the $s$ integral into 
two sub-ranges $0<s<k_1 \sim k_3$ 
and $s>k_1 \sim k_3 $. 
Once again we approximate the integrands by assuming $s\ll k_1 \sim k_3$ and $s\gg k_1 \sim k_3$ for the respective sub-ranges 
to give
\EQ
\psi_{123} = \frac{1}{(4\pi p_\gamma)^3} ~ \mathcal I = \left( \frac{3}{4\pi\rho_0} \right)^3 ~ \mathcal I
\EN
where 
\EQA
\mathcal I &=& \int d^3s M(\vert \kk_1 + \ssvec \vert) M(s) M(\vert \ssvec   -\kk_3 \vert)  \nonumber \\
&\quad& \times \left[ -8 + \bar{\beta}^2 + 9 \left( \mu^2 +2 \gamma^2  \right) + 6 \mu \bar{\beta} \gamma + 3 \bar{\beta}^2 \gamma^2 \right. \nonumber \\
&\quad& - \left. 9 \gamma^2 \left( 3\mu^2 + \gamma^2 \right) - 18 \mu \bar{\beta} \gamma^3 + 27 \mu^2 \gamma^4 \right]  \nonumber \\
\ENA
Performing the integrals for all the terms, with $n \to -3$, we find that there
is considerable though incomplete cancellation between the 10 terms 
\EQA
\mathcal I &\simeq& 2 \pi A^3 \frac{k_1^{2n+3}k_3^n}{n+3} \times \nonumber \\ 
&&\left[ -16 +\frac{2}{3} + 6 + 36 + 4 + 2 - 18 - 18 - 12 +18  \right] \nonumber \\
&\simeq& 2 \pi A^3  \frac{k_1^{2n+3}k_3^n}{n+3} \left[ \frac{8}{3} \right] 
\ENA
We draw attention to how the full evaluation gives a result for $\psi_{123}$
that is of opposite sign and one-sixth magnitude (8/3) of the value if
we only consider the constant term (-16) for this squeezed collinear
configuration. The sign of the mode-coupling integral is important as the
bispectrum is not a positive-definite quantity (unlike the power spectrum) and
the observed limits on the bispectrum may also be asymmetric about zero. Then
the Case II - squeezed collinear mode-coupling integral is
\EQ
\psi_{123} = \left[ \frac{8}{3} \right] ~2 \left( 4 \right)^3 \frac{\pi^7}{k_G^6}
(n+3)^2 \left(\frac{k_1}{k_G}\right)^{2n+3} \left(\frac{k_3}{k_G}\right)^n V_A^6.
\label{psi_II}
\EN
As the n-dependence of $\psi_{123}$ is identical for both Case I and Case II we can write 
\EQ
\psi_{123} =  \mathcal K \left[ ~2~ \left( 4 \right)^3 \frac{\pi^7}{k_G^6}
(n+3)^2 \left(\frac{k_1}{k_G}\right)^{2n+3} \left(\frac{k_3}{k_G}\right)^n V_A^6.\right] 
\label{psi_both}
\EN
where 
\EQ
\mathcal K = \left\{
\begin{array}{ll}
2 \, m & \text{   Case I} \\
\frac{8}{3} & \text{   Case II} \\
\end{array}
\right.
\EN

\section{Passive Scalar CMB Bispectrum}

Using \Eq{psi_both} in \Eq{bispec} we can evaluate the CMB bispectrum for the passive scalar mode for both Case I and II. 
The algebraic steps are the same as those for the compensated scalar mode in SS09.
We express the delta function present in $\zeta_{_{123}}$ in its integral form $\delta(\kk) = (1/(2\pi)^3)\int d^3x 
\exp(i\kk\cdot\xx)$, use the spherical wave expansion of the exponential terms, substitute it into \Eq{bispec}, 
and integrate over the angular parts of $(\kk_1,\kk_2,\kk_3,\xx)$. This algebra is also very similar to that in the 
calculation of the primordial bispectrum \cite{FS07}.
Then it becomes possible to write the bispectrum $B^{m_{_1}m_{_2}m_{_3}}_{l_{_1}l_{_2}l_{_3}}$,
in terms of a reduced bispectrum $b_{l_{_1}l_{_2}l_{_3}}$ (also referred to as the 
Komatsu-Spergel estimator \cite{komatsu_spergel01}) as
\EQ
B^{m_{_1}m_{_2}m_{_3}}_{l_{_1}l_{_2}l_{_3}} =
{\cal G}_{m_{_1}m_{_2}m_{_3}}^{l_{_1}l_{_2}l_{_3}}
~ b_{l_{_1}l_{_2}l_{_3}}
\label{gaunt1}
\EN
where
\EQA
b_{l_{_1}l_{_2}l_{_3}} &=&
\left(\frac{{\cal R}_p}{\pi^2}\right)^3 \int x^2 dx
\nonumber \\
&\quad& \times \prod_{i=1}^3
\int k_i^2dk_i ~ j_{_{l_i}}(k_{_i}x) ~ j_{_{l_i}}(k_{_i}D^*)
~ \psi_{123}
\label{redbispec}
\ENA
and we have introduced the Gaunt integral
\EQ
{\cal G}_{m_{_1}m_{_2}m_{_3}}^{l_{_1}l_{_2}l_{_3}}
= \int d\Omega ~ Y_{l_1m_1}Y_{l_2m_2}Y_{l_3m_3} .
\label{gaunt}
\EN
We substitute \Eq{psi_both} into \Eq{redbispec} for the reduced bispectrum.
The integrals over $k_2$ can  be immediately done using the spherical Bessel function closure relation
\EQ
\int k_2^2 dk_2 j_{l_2}(k_2x) j_{l_2}(k_2D^*) = (\pi/2x^2) \delta(x - D^*).
\EN
and the delta function makes the $x$-integral trivial. We are then left with integrals over $k_1$ and $k_3$ given by
\EQA
b_{l_{_1}l_{_2}l_{_3}} &=& 
\mathcal K \left(\frac{{\cal R}_p}{\pi^2}\right)^3
 \frac{\pi}{2} \left[ ~2~ \left( 4 \right)^3 \pi^7 (n+3)^2 V_A^6.\right] \nonumber \\
&\quad& \times \left[\int \frac{dk_3}{k_3} j_{_{l_3}}^2(k_3D^*)
\left(\frac{k_3}{k_G}\right)^{n+3}
\right]
\nonumber \\
&\quad& \times
\left[\int \frac{dk_1}{k_1}j_{_{l_1}}^2(k_1D^*)
\left(\frac{k_1}{k_G}\right)^{2(n+3)}
\right]
\label{bispec_k_integral}
\ENA
The $k_3$ and $k_1$ integrals are similar in form to the usual Sachs-Wolfe term
(for power-law magnetic spectra) and can be evaluated analytically for any n
(Eq. 6.574.2 of \cite{Gradshteyn6}) in terms of Gamma functions. For $n \to -3$
we have
\EQA
b_{l_{_1}l_{_2}l_{_3}} &=& 
\mathcal K \left(\frac{{\cal R}_p}{\pi^2}\right)^3
 \frac{\pi}{2} \left[ ~2~ \left( 4 \right)^3 \pi^7 (n+3)^2 V_A^6.\right] \nonumber \\
&\quad& \times \frac{1}{2 l_3 (l_3+1)} \frac{1}{2 l_1 (l_1+1)}
\label{bispec_l1_l3}
\ENA
or
\EQ
l_1 (l_1+1) l_3 (l_3+1) b_{l_{_1}l_{_2}l_{_3}} =
\mathcal K ~{{\cal R}_p}^3 \left( \frac{\pi}{2}\right)^2 \left( 4 \right)^3 (n+3)^2 V_A^6
\label{bispec_symbolic}
\EN
where a factor of $1/(D^* k_G)^{3(n+3)}$ that also appears has been approximated to unity 
for the case $n \to -3$ of a scale-invariant magnetic field index.

For \underline{Case I} - The bispectrum considering only constant angular terms
\EQA
l_1 (l_1+1) l_3 (l_3+1) b_{l_{_1}l_{_2}l_{_3}} &\approx& 
\left( -1.35 \times 10^{-16} \right) \left( \frac{3m}{4}\right)  \nonumber \\
&\quad& \times \left( \frac{n+3}{0.2}\right)^2 \left( \frac{B_{-9}}{3}\right)^3
\label{bispec_caseI}
\ENA
where we have used $\tau_B \approx 10^{14}$ GeV corresponding to a possible
inflationary epoch for magnetic field generation. This evaluates in different configurations to
\EQA
&&l_1 (l_1+1) l_3 (l_3+1) b_{l_{_1}l_{_2}l_{_3}} \approx \nonumber \\
 && \left(  10^{-16} \right) \left( \frac{n+3}{0.2}\right)^2 \left( \frac{B_{-9}}{3}\right)^3 \times
\left\{
\begin{array}{ll}
0 & \text{local isosceles} \\
5.7 & \text{equilateral} \\
8.1 & \text{squeezed collinear} \\
9.2 & \text{midpoint collinear} \\
\end{array}
\right\} \nonumber \\
\label{bispec_caseI_values}
\ENA
We see that the constant-term only evaluation gives a bispectrum of order $6-9
\times 10^{-16}$ independent of the exact configuration. The exception is the
local isosceles case (where the constant term sum happens to vanish) but if a
full evaluation over all angular terms is done, it yields a non-zero bispectrum.

For \underline{Case II} - The squeezed collinear bispectrum considering all angular terms
\EQA
l_1 (l_1+1) l_3 (l_3+1) b_{l_{_1}l_{_2}l_{_3}} &\approx& 
\left( -1.35 \times 10^{-16} \right)   \nonumber \\
&\quad& \times \left( \frac{n+3}{0.2}\right)^2 \left( \frac{B_{-9}}{3}\right)^3
\label{bispec_caseII}
\ENA
The full evaluation of all angular terms for the squeezed collinear case
produces two important changes (compared to squeezed collinear in Case I
evaluation): the sign of the bispectrum has changed to negative (angular terms
in mode-coupling contribute with different signs and change the total sign) and
it's value had diminished by factor of 6. 

Note that the values of the reduced bispectrum given by Equations (\ref{bispec_caseI_values}) and (\ref{bispec_caseII}) are 
$\sim 10^6$ times larger than the values obtained for the compensated scalar mode in SS09. This is essentially due 
to the large value of ${\cal A}^{3}$ as discussed in section 4. 

A numerical evaluation of the mode-coupling integral involves an integrable singularity at $\ssvec = \kk_3$ and $\ssvec = -\kk_1$. 
The singularity is integrable (even without a large-scale cutoff) provided we consider the mode-coupling integral over $s$ 
together with the $k_1$ and $k_3$ integrals. We defer this numerical investigation for later work and proceed with our 
analytic results.

%*=*=*=*=*=*=*=*=*=*=*
\section{Constraint on Primordial Magnetic Field}

In order to put constraints on the primordial magnetic field,
we compare our magnetic reduced bispectrum with the reduced bispectrum that
arises from non-linear terms in the gravitational potential. These have a value
given by 
\EQ
l_1(l_1+1)l_3(l_3+1) b_{l_{_1}l_{_2}l_{_3}} \sim 4 \times 10^{-18} f_{NL}
\label{inflationary_bispec}
\EN
characterized by $f_{NL}$ (\cite{riotto08}). 

The reduced bispectrum for the magnetic compensated scalar mode \cite{SS09,caprini09} was a factor of a few 
times $10^4$ smaller than the standard inflationary signal (for models with $f_{NL} \sim 1 $). However, we find that 
the magnitude of the corresponding 
reduced bispectrum for the magnetic passive scalar mode, calculated above, is $\sim 30$ times larger (Case II) or even up to 
$\sim 200$ times larger (Case I) than the inflationary signal. Both these magnetic bispectra values are for field strength of 
3 nG and $n=-2.8$, close to scale-invariant spectrum.

Conversely, we can put upper limits on the primordial magnetic field by equating the magnetic bispectrum to the observed 
upper limit for the inflationary bispectrum. This is evaluated using the latest WMAP7 limits (95 \% CL) on non-Gaussianity 
in the observed CMB \cite{komatsu_wmap7}, $-10 < f_{NL}^{loc} < 74$, taking the appropriate side of these limits for the 
different signs of bispectrum. Since we have considered only the Sachs-Wolfe contribution to magnetic field-induced 
temperature anisotropies, our magnetic bispectra will be accurate only on large scales. 
However, the $f_{NL}$ we are using for comparison to data comes from WMAP data that has a maximum multipole of $l \sim 750$ 
and therefore the bispectra comparison is not exact. Yet our $B_0$ limits are expected to be fairly robust as $B_0$ 
depends very weakly ($ B_0 \propto f_{NL}^{1/6}$) on $f_{NL}$ as also discussed in SS09.
We obtain the following upper limits for the magnetic field in different cases:\\

\underline{Case I}: For any bispectrum configuration considering only $s$-independent terms
\EQ
B_0 < 3 \text{ nG} 
\EN
For individual configurations the magnetic field limits are: equilateral (2.7 nG), squeezed collinear (2.5 nG) 
and midpoint collinear (2.5 nG), all for $\tau_B \approx 10^{14}$ GeV corresponding to the inflationary epoch.\\

\underline{Case II}: For the squeezed collinear bispectrum configuration, including all angular terms, we obtain an upper limit
\EQ
B_0 < 2 \text{ nG}
\EN
where this magnetic field limit value (2.4 nG) is derived using $\tau_B \approx 10^{14}$ GeV and 
$ -10 <  f_{NL}^{loc} $.

It is currently unclear what would be an appropriate epoch to adopt for
the generation of a large-scale primordial cosmic magnetic field \cite{widrow02,
TW88,ratra92,giovannini_string07,martin_yoko08,kandu_AN10}. We recall that
via the logarithmic factor in Eq. (\ref{zeta}), there is a weak dependence of
$B_0$ on $\tau_B$, the magnetic field generation epoch. We find a 2-4 nG
variation in the $B_0$ upper limit when $\tau_B$ is varied between the inflationary and
electroweak epochs $\tau_B \approx 10^{14} \to 10^{3} $ GeV, taken as the earliest and 
latest possible epochs of magnetic field generation. 
However, as the electroweak transition is not expected to give a scale-invariant 
magnetic field spectrum, we caution that the choice of electroweak epoch may be too 
late to generate a field with $n \to -3$. Hence,
the variation in the magnetic field upper limit quoted above may be an overestimate.

%*=*=*=*=*=*=*=*=*=*=*
\section{Conclusion}
We have calculated here the CMB bispectrum, on large angular scales,
due to the passive scalar mode. This mode is sourced by
the magnetic scalar anisotropic stress before neutrino decoupling.
The CMB bispectrum due to the passive scalar mode
is more than two orders of magnitude larger
than the bispectrum due to the primary inflation-induced scalar mode,
for $f_{NL} \sim 1$ and $B_0 \sim$ 3 nG.
It is also a factor of $10^6$ times greater
than that previously calculated for the compensated scalar
mode \cite{SS09,caprini09}.
This is the first calculation of a type of magnetic
contribution to CMB bispectrum
that can clearly dominate over the primary bispectrum at large scales.
In the CMB power spectrum, by contrast, the passive scalar mode signal is highly sub-dominant
to the inflationary signal \cite{SL09}.
Our work thus shows that the magnetically induced signals,
being intrinsically non-Gaussian, could be more easily distinguished from the
primary CMB anisotropies, when one considers the bispectrum rather
than the power spectrum.

Using the WMAP7 limits for $f_{NL}$ we have placed an upper limit on the
strength of the primordial magnetic field $B_0 < 2$ nG which is more than an
order of magnitude stronger than the limit from the compensated scalar mode (35
nG) obtained in SS09. The passive mode limit is only weakly
(logarithmically) dependent on the epoch of magnetic field generation $\tau_B$. 
We note that for scale-invariant magnetic spectra $n \to -3$, the bispectrum and
the derived magnetic field limit are independent of both the assumed
cosmological model as well as our choice of smoothing scale $k_G$ for the
stochastic magnetic field $B_0$. 

The primordial magnetic field limit is expected to improve if the ISW effect and
full scalar anisotropies are taken into account. We also expect stronger
constraints to follow from the tensor and vector mode bispectra that dominate at
low and high $l$, respectively. Better observational limits on CMB
non-Gaussianity from PLANCK data are expected to further tighten
primordial magnetic field limits.

\section*{Acknowledgements}
PT would like to acknowledge the assistance from Sri Venkateswara College, University of Delhi, in pursuing this work. 
PT and TRS would like to acknowledge the IUCAA Associateship Program as well as 
the facilities at the IUCAA Resource Center, University of Delhi.


\begin{thebibliography}{99}

\bibitem{beck09}
R. Beck, Astrophys. Space Sci. Trans. \textbf{5}, 43 (2009).

\bibitem{vogt_ensslin03}
C. Vogt and T. A. En\ss lin, Astron. \& Astrophys. \textbf{412}, 373 (2003). 

\bibitem{BS05}
A. Brandenburg and K. Subramanian, Phys. Rep. \textbf{417}, 1 (2005).

\bibitem{kulsrud_zweibel08}
R. M. Kulsrud and E. G. Zweibel, Rep. Prog. Phys. \textbf{71}, 4, 046901 (2008).

\bibitem{beck_ARAA}
R. Beck, A. Brandenburg, D. Moss, A. Shukurov and D. Sokoloff, Annu. Rev. Astron. Astrophys. \textbf{34}, 155 (1996).

\bibitem{bernet08}
M. L. Bernet, F. Miniati, S. J. Lilly, P. P. Kronberg and M. Dessauges-Zavadsky, Nature \textbf{454}, 7202, 302 (2008).

\bibitem{kronberg08}
P. P. Kronberg, M. L. Bernet, F. Miniati, S. J. Lilly, M. B. Short, D. M. Higdon, Astrophys. J. \textbf{676}, 70 (2008).

\bibitem{kim89}
K.-T. Kim, P. P. Kronberg, G. Giovannini and T. Venturi, Nature \textbf{341}, 720 (1989).

\bibitem{ando10} 
A. Ando and A Kusenko, ArXiv e-prints (2010), arXiv:1005.1924v2  [astro-ph.HE].

\bibitem{neronov10} 
A. Neronov and I Vovk, Science \textbf{328}, 73 (2010).

\bibitem{kandu_AN06}
K. Subramanian, Astron. Nach. \textbf{327}, 333 (2006).

\bibitem{durrer_NAR07}
R. Durrer, New Astron. Rev. \textbf{51}, 275 (2007).

%=============
%\bibitem{Bcmb}

\bibitem{barrow97}
J. D. Barrow, P. G. Ferreira and J. Silk, Phys. Rev. Lett. \textbf{78}, 3610 (1997).

\bibitem{SB98PRL}
K. Subramanian and J. D. Barrow, Phys. Rev. Lett. \textbf{81}, 3575 (1998).

\bibitem{durrer00}
R. Durrer, P. G. Ferreira, T. Kahniashvili, Phys. Rev. D \textbf{61}, 043001 (2000).

\bibitem{trsks01}
T. R. Seshadri and K. Subramanian, Phys. Rev. Lett. \textbf{87}, 101301 (2001).

\bibitem{mack02}
A. Mack, T. Kahniashvili and A. Kosowsky, Phys. Rev. D \textbf{65}, 123004 (2002).

\bibitem{lewis04}
A. Lewis, Phys. Rev. D \textbf{70}, 043011 (2004).

\bibitem{SS05}
T. R. Seshadri and K. Subramanian, Phys. Rev. D \textbf{72}, 023004 (2005).

\bibitem{gopal_sethi05}
R. Gopal and S. K. Sethi, Phys. Rev. D \textbf{72}, 103003 (2005).

\bibitem{kahn05}
T. Kahniashvili and B. Ratra, Phys. Rev. D \textbf{71}, 103006 (2005).

\bibitem{giovannini06}
M. Giovannini, Phys. Rev. D \textbf{74}, 063002 (2006).

\bibitem{giovan_kunze08}
M. Giovannini and K. E. Kunze, Phys. Rev. D \textbf{77}, 063003 (2008).

\bibitem{yam08}
D. G. Yamazaki, K. Ichiki, T. Kajino, and G. J. Mathews, Phys. Rev. D \textbf{77}, 043005 (2008).

\bibitem{finelli08}
F. Finelli, F. Paci and D. Paoletti, Phys. Rev. D \textbf{78}, 023510 (2008).

\bibitem{pao09}
D. Paoletti, F. Finelli and F. Paci, Mon. Not. R. Astron. Soc. \textbf{396}, 1, 523 (2008).

\bibitem{giovan09}
M. Giovannini, Phys. Rev. D \textbf{79}, 121302 (2009).
%================
% Bcmb

\bibitem{yam10} 
D. G. Yamazaki, K. Ichiki, T. Kajino and G. Mathews, Phys. Rev. D \textbf{81}, 023008 (2010). 

\bibitem{pao10} 
D. Paoletti and F. Finelli, ArXiv e-prints (2010), arXiv:1005.0148 [astro-ph.CO].

\bibitem{SL10}
J. R. Shaw and A. Lewis, ArXiv e-prints (2010), arXiv:1006.4242v1 [astro-ph.CO].

\bibitem{widrow02} 
L. M. Widrow, Rev. Mod. Phys. \textbf{74}, 775 (2002).

%===================================
\bibitem{moffatt}
H. K. Moffatt, {\it Magnetic Field Generation in Electrically Conducting Fluids} 
(Cambridge University Press, Cambridge, 1978).

\bibitem{krause}
F. Krause, K.-H. R\"adler,
{\it Mean-field Magneto\-hydrodynamics and Dynamo Theory} 
(Pergamon Press, Oxford, 1980).

\bibitem{ruzmaikin}
A. A. Ruzmaikin, A. Shukurov and D. Sokoloff, {\it
Magnetic Fields of Galaxies} (Kluwer, Dordrecht, 1988).
%====================================


%===================
\bibitem{TW88}
M. Turner, L. M. Widrow, Phys. Rev. D \textbf{37}, 2743 (1988).

\bibitem{ratra92}
B. Ratra, Astrophys. J. \textbf{391}, L1 (1992).

\bibitem{giovannini_string07}
M. Giovannini, {\it String theory and fundamental interactions},
edited by M. Gasperini and J. Maharana, Lecture Notes in Physics (Springer,
Berlin/Heidelberg, 2007) (arXiv:astro-ph/0612378).

\bibitem{martin_yoko08}
J. Martin and J. Yokoyama, Journal of Cosmology and Astroparticle Physics \textbf{01}, 025 (2008).

\bibitem{kandu_AN10}
K. Subramanian, Astron. Nachr. \textbf{331}, 1, 110 (2010).

%====================

\bibitem{SL09}
J. R. Shaw and A. Lewis, Phys. Rev. D \textbf{81}, 043517 (2010).

\bibitem{brown10}
I. A. Brown, ArXiv e-prints (2010), arXiv:1005.2982v1 [astro-ph.CO].

%==================================
% homB
\bibitem{chen04}
G. Chen, P. Mukherjee, T. Kahniashvili, B. Ratra and Y. Wang, Astrophys. J. \textbf{611}, 655 (2004).

\bibitem{naselksy04}
P. D. Naselsky, L-Y. Chiang, P. Olesen and O. V. Verkhodanov, Astrophys. J. \textbf{615}, 45 (2004).

\bibitem{bernui08}
A. Bernui and W.S. Hipolito-Ricaldi, Mon. Not. R. Astron. Soc. \textbf{389}, 1453 (2008).

\bibitem{kahn08}
T. Kahniashvili, G. Lavrelashvili and B. Ratra, Phys. Rev. D \textbf{78}, 063012 (2008).
% homB
%====================================

\bibitem{wang_kamion00}
L. Wang and M. Kamionkowski, Phys. Rev. D \textbf{61}, 063504 (2000).

\bibitem{komatsu_spergel01}
E. Komatsu and D. N. Spergel, Phys. Rev. D \textbf{63}, 063002 (2001).

\bibitem{bartolo04}
N. Bartolo, E. Komatsu, S. Matarrese and A. Riotto, Phys. Rep. \textbf{402}, 3, 103266 (2004).

\bibitem{FS07}
J. R. Fergusson and E. P. S. Shellard, Phys. Rev. D \textbf{76}, 083523 (2007).

\bibitem{riotto08}
A. Riotto, {\it The Quest for Non-Gaussianity}, Lect. Notes Phys. \textbf{738}, 305 (2008).

\bibitem{yadav_wandelt10}
A. P. S. Yadav and B. D. Wandelt, Advances in Astronomy \textbf{2010}, 565248 (2010) (doi:10.1155/2010/565248).

\bibitem{SS09}
T. R. Seshadri and K. Subramanian, Phys. Rev. Lett. \textbf{103}, 081303 (2009).

\bibitem{caprini09}
C. Caprini, F. Finelli, D. Paoletti and A. Riotto, Journal of Cosmology and Astroparticle Physics \textbf{06}, 021 (2009). 

\bibitem{cai10}
R.-G. Cai, B. Hu and H.-B. Zhang, Journal of Cosmology and Astroparticle Physics \textbf{08}, 025 (2010).

\bibitem{BC05}
I. Brown, and R. Crittenden, Phys. Rev. D \textbf{72}, 063002 (2005).

\bibitem{kojima10}
K. Kojima, T. Kajino and G. J. Mathews, Journal of Cosmology and Astroparticle Physics \textbf{02}, 018 (2010).

\bibitem{bonvin10}
C. Bonvin and C. Caprini, Journal of Cosmology and Astroparticle Physics \textbf{05}, 022 (2010).

\bibitem{kunze10}
K. E. Kunze, ArXiv e-prints (2010), arXiv:1007.3163v1 [astro-ph.CO].

\bibitem{jedam98} 
K. Jedamzik, V. Katalinic and A. Olinto, Phys. Rev. D {\bf 57}, 3264 (1998). 

\bibitem{SB98} 
K. Subramanian and J. D. Barrow, Phys.Rev. D {\bf 58} 083502 (1998).

\bibitem{SSB03}
K. Subramanian, T. R. Seshadri and J. D. Barrow, Mon. Not. Roy. Astr. Soc. \textbf{344}, L31 (2003).

\bibitem{cap02} 
C. Caprini, R. Durrer, Phys. Rev. D \textbf{65}, 023517 (2002).

\bibitem{giovannini10_spectator}
M. Giovannini, Phys. Rev. D \textbf{81}, 127302 (2010).

\bibitem{lythliddle}
D. H. Lyth and A. R. Liddle, \textit{The Primordial Density Perturbation} (Cambridge University Press, Cambridge U.K., 2009).

\bibitem{giovannini07_PMC}
M. Giovannini, PMC Physics A, \textbf{1}:5 (2007) (doi:10.1186/1754-0410-1-5).

\bibitem{Gradshteyn6}
I. S. Gradshteyn and I. M. Ryzhik, \textit{Table of Integrals, Series and Products} 
(Academic Press, New York, U.S.A. and London U.K., 6th edition, 2000).

\bibitem{komatsu_wmap7}
E. Komatsu et al.,ArXiv e-prints (2010), arXiv1001.4538 [astro-ph.CO].

\end{thebibliography}
\end{document}